# Declarative vs Rule-based Control for Flocking Dynamics

Usama Mehmood
Department of Computer Science,
Stony Brook University, USA

Nicola Paoletti
Department of Computer Science,
Stony Brook University, USA

Dung Phan
Department of Computer Science,
Stony Brook University, USA

Radu Grosu
Cyber-Physical Systems Group,
Technische Universitat Wien, Austria

Shan Lin
Department of Electrical and
Computer Engineering, Stony Brook
University, USA

Scott D. Stoller
Department of Computer Science,
Stony Brook University, USA

Ashish Tiwari
SRI International, USA

Junxing Yang
Department of Computer Science,
Stony Brook University, USA

Scott A. Smolka
Department of Computer Science,
Stony Brook University, USA

## ABSTRACT

The popularity of rule-based flocking models, such as Reynolds' classic flocking model, raises the question of whether more declarative flocking models are possible. This question is motivated by the observation that declarative models are generally simpler and easier to design, understand, and analyze than operational models. We introduce a very simple control law for flocking based on a cost function capturing cohesion (agents want to stay together) and separation (agents do not want to get too close). We refer to it as *d*eclarative *f*locking (DF). We use model-predictive control (MPC) to define controllers for DF in centralized and distributed settings. A thorough performance comparison of our declarative flocking with Reynolds' classic flocking model, and with more recent flocking models that use MPC with a cost function based on lattice structures, demonstrate that DF-MPC yields the best cohesion and least fragmentation, and maintains a surprisingly good level of geometric regularity while still producing natural flock shapes similar to those produced by Reynolds' model. We also show that DF-MPC has high resilience to sensor noise.

## 1 INTRODUCTION

Flocking is a collective behavior exhibited by a large number of interacting agents possessing a common group objective [7]. The term is most commonly associated with birds, and more recently, drones. Examples include foraging for food, executing a predator-avoidance maneuver, and engaging in migratory behavior.

With the introduction of Reynolds' model [12, 13], *rule-based control* became the norm in the flocking community. Specifically, in this model, at each time-step, each agent executes a control law given in terms of the weighted sum of three competing forces to determine its next acceleration. Each of these forces has its own rule: *separation* (keep a safe distance away from your neighbors), *cohesion* (move towards the centroid of your neighbors), and *alignment* (steer toward the average heading of your neighbors). As the descriptions suggest, these rules are executed by each agent in a distributed environment with limited-range sensing and no communication.

The popularity of Reynolds' model and its many variants raises the question: Is there a more abstract *declarative* form of control for flocking? This question is important because declarative models are generally simpler and easier to design, understand, and analyze than operational models. This is analogous to declarative programs (e.g., functional programs and logic programs) being easier to write and verify than imperative programs.

We show that the answer to this question is indeed positive by providing a very simple control law for flocking based on a *cost function* comprising two main terms: *cohesion* (the average squared distance between all pairs of agents) and *separation* (a sum of inverse squared distances, except this time between pairs of agents within each other's sensing range). That is it. For example, no term representing velocity alignment is needed. The cost function specifies what we want as the goal, and is hence declarative. In contrast, the update rules in Reynolds' model aim to achieve an implicit goal and hence are operational. Executing declarative control amounts to finding the right balance between attracting and repelling forces between agents. We refer to this approach as *Declarative Flocking* (DF). We use MPC (model-predictive control) to define controllers for DF, and refer to this approach as *DF-MPC*. We define a centralized version of DF-MPC, which requires communication, and a distributed version, which does not.

Previous MPCs for flocking exist, e.g., [16–18]. Most of these MPCs are designed to conform to the $\alpha$-lattice model of flocking proposed in [7]. $\alpha$-lattices impose a highly regular structure on flocks: all neighboring agents are distance $d$ apart, for a specified constant $d$. This kind of structure is seen in some settings, such as beehives, but is not expected in many other natural and engineered settings, and it is not imposed by Reynolds' model.

In this paper, we show, via a thorough performance evaluation, how centralized and distributed DF-MPC compare to Reynolds' rule-based approach [12, 13], Olfati-Saber's potential-based approach [7], a variant of Zhan and Li's centralized lattice-based MPC approach [15, 16], and Zhang et al.'s distributed lattice-based MPC approach [17]. We consider performance measures that capture multiple dimensions of flocking behavior: number of sub-flocks (flock fragmentation), maximum sub-flock diameter (cohesion), velocity convergence, and a new parameter-free measure of the geometric regularity of the formation.

Our experimental results demonstrate that DF-MPC yields the best cohesion and least fragmentation, and produces natural flock shapes like those produced by Reynolds' model. Also, distributed DF-MPC maintains a surprisingly good level of geometric regularity. We also analyze the resiliency of DF-MPC and the lattice-based MPC approaches by considering the impact of sensor noise. Our results demonstrate a remarkably high level of resiliency on the part of DF-MPC in comparison with these other approaches.

The rest of the paper is organized as follows. Section 2 presents the rule-based, potential-based, and lattice-based MPC approaches mentioned above. Section 3 defines our declarative flocking approach. Section 4 defines our performance measures for flocking models. Section 5 presents our experimental results and performance evaluation. Section 6 discusses related work. Finally, Section 7 offers our concluding remarks and directions for future work.

## 2 MODELS OF FLOCKING BEHAVIOR

We consider a set of dynamic *agents* $\mathcal{B} = \{1, \ldots, n\}$ that move according to the following discrete-time equation of motion:

$$x_i(k+1) = x_i(k) + dT \cdot v_i(k), \; v_i(k) \in V \quad (1)$$

$$v_i(k+1) = v_i(k) + dT \cdot a_i(k), \; a_i(k) \in A, \quad (2)$$

where $x_i(k), v_i(k), a_i(k) \in \mathbb{R}^m$ are respectively position, velocity and acceleration of agent $i \in \mathcal{B}$ in the $m$-dimensional space at step $k$, and $dT \in \mathbb{R}^+$ is the time step. We consider physical constraints on velocities and accelerations, described by the sets $V$ and $A$, respectively, which are defined by $V = \{v \mid |v| \leq \bar{v}\}$ and $A = \{a \mid |a| \leq \bar{a}\}$, where $\bar{v}$ and $\bar{a}$ limit the allowed magnitude of the velocity and acceleration vectors, respectively.

In most flocking models, agents update their motion by changing their acceleration. In this sense, $a_i(k)$ represents the control input for agent $i$.

The *configuration* of all agents is described by the vector $\mathbf{x}(k) = [x_1^T(k) \ldots x_n^T(k)]^T \in \mathbb{R}^{m \cdot n}$. Let $\mathbf{v}(k) = [v_1^T(k) \ldots v_n^T(k)]^T \in \mathbb{R}^{m \cdot n}$, and $\mathbf{a}(k) = [a_1^T(k) \ldots a_n^T(k)]^T \in \mathbb{R}^{m \cdot n}$. Then the equation of motion for all agents can be expressed as

$$\mathbf{x}(k+1) = \mathbf{x}(k) + dT \cdot \mathbf{v}(k), \quad (3)$$

$$\mathbf{v}(k+1) = \mathbf{v}(k) + dT \cdot \mathbf{a}(k), \quad (4)$$

The local neighborhood of agent $i$ is defined by the set of other agents, called *neighbors*, within a given distance from $i$, mimicking the agent's visibility sphere. For an *interaction radius* $r > 0$ and configuration $\mathbf{x}$, the set of *spatial neighbors* of agent $i$, $N_i(\mathbf{x}) \subseteq \mathcal{B}$, is given by:

$$N_i(\mathbf{x}) = \{j \in \mathcal{B} \mid j \neq i \wedge \|x_i - x_j\| < r\}, \quad (5)$$

where $\|\cdot\|$ denotes the Euclidean norm.

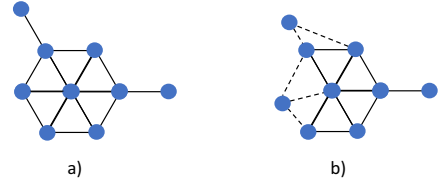

**Figure 1: Examples of $\alpha$-lattice a) and quasi $\alpha$-lattice b). Solid lines connect agents in the same neighborhood that have distance $d$. Dashed lines connect those with have distance $d \pm \epsilon$ for $\epsilon \leq \delta$ (the tolerance).**

For configuration $\mathbf{x} \in \mathbb{R}^{m \cdot n}$, we define the associated *proximity net* $G(\mathbf{x}) = (\mathcal{B}, \mathcal{E}(\mathbf{x}))$ as the graph that connects agents within their interaction radius:

$$\mathcal{E}(\mathbf{x}) = \{(i, j) \in \mathcal{B} \times \mathcal{B} \mid \|x_i - x_j\| < r, i \neq j\}, \quad (6)$$

To capture the regular geometry of flocks, Olfati-Saber introduced the notions of $\alpha$-*lattices*, i.e. configurations where each agent is equally distant from its neighbors, and *quasi $\alpha$-lattices*, i.e. configurations that are $\alpha$-lattices modulo a small error in the distances [7]. The *scale* parameter $d$ defines the ideal inter-agent distance.

*Definition 2.1 ($\alpha$-lattice [7]).* A configuration $\mathbf{x} \in \mathbb{R}^{m \cdot n}$ is called $\alpha$-lattice if for all $i \in \mathcal{B}$ and all $j \in N_i(\mathbf{x})$, $\|x_i - x_j\| = d$, where $d \in \mathbb{R}^+$ is the scale of the $\alpha$-lattice. For tolerance $\delta \in \mathbb{R}^+$, a configuration $\mathbf{x} \in \mathbb{R}^{m \cdot n}$ is called a *quasi $\alpha$-lattice* if for all $i \in \mathcal{B}$ and all $j \in N_i(\mathbf{x})$, $|\|x_i - x_j\| - d| \leq \delta$.

### 2.1 Sensing noise

We extend the classical equations of motion, Eqs. (1)–(2), with *sensing noise* affecting how each agent perceives positions and velocities of its neighbors. Existing work has put little focus on flocking dynamics subject to noise, which is unfortunately unavoidable in realistic natural and engineered flocks.

For actual positions $\mathbf{x}(k)$ and velocities $\mathbf{v}(k)$ at step $k$, let $\tilde{\mathbf{x}}(k)$ and $\tilde{\mathbf{v}}(k)$ denote their noisy counterparts sensed by a generic agent, defined by:

$$\tilde{\mathbf{x}}(k) = \mathbf{x}(k) + \mathbf{nx}(k) \text{ and } \tilde{\mathbf{v}}(k) = \mathbf{v}(k) + \mathbf{nv}(k), \quad (7)$$

where $\mathbf{nx}(k)$ and $\mathbf{nv}(k)$ in $\mathbb{R}^{m \cdot n}$ are vectors of independent and identically distributed (i.i.d.) random variables. The position noise $\mathbf{nx}(k)$ and velocity noise $\mathbf{nv}(k)$ are distributed according to Gaussian distributions with mean 0 and standard deviation $\sigma_x$ and $\sigma_v$, respectively. We stress the dependency on $k$ because noise variables are independent across time steps.

In centralized flocking algorithms, where agent decisions are computed by a single controller with information about the whole population, we use Eq. 7 to define noisy measurements. In distributed algorithms, sensing noise is independent for each agent. We denote the noisy measurement of agent $i$ by $\tilde{\mathbf{x}}^{\triangleright i}(k)$ and $\tilde{\mathbf{v}}^{\triangleright i}(k)$, where positions and velocities are noisy for all agents except agent $i$

$$\tilde{\mathbf{x}}^{\triangleright i}(k) = [\tilde{x}_1^T(k) \ldots x_i^T(k) \ldots \tilde{x}_n^T(k)]^T \text{ and} \quad (8)$$

$$\tilde{\mathbf{v}}^{\triangleright i}(k) = [\tilde{v}_1^T(k) \ldots v_i^T(k) \ldots \tilde{v}_n^T(k)]^T, \quad (9)$$

with $\tilde{x}_1(k), \ldots, \tilde{x}_n(k)$ and $\tilde{v}_1(k), \ldots, \tilde{v}_n(k)$ defined as per (7); implicitly, for each agent $i$ and each other agent $j$, the noise distribution is sampled independently to compute the $\tilde{x}_j^T(k)$ component of $\tilde{\mathbf{x}}^{\triangleright i}(k)$.

## 2.2 Reynolds' rule-based model

In Reynolds' rule-based distributed model [12, 13], agents follow simple rules to compute their accelerations from the positions and velocities of their neighbors. The rules are illustrated in Figure 2. They do not explicitly specify the desired flocking formation as an objective; rather, flocking emerges from the interaction rules.

Specifically, each agent $i \in \mathcal{B}$ updates its acceleration $a_i(k)$ at step $k$ by considering the following three components (adapted to include sensing noise):

(1) *Alignment*: agents match their velocities with the average velocity of nearby agents.

$$a_i^{al}(k) = w_{al} \cdot \left( \left( \frac{1}{|N_i(\tilde{\mathbf{x}}^{\triangleright i}(k))|} \cdot \sum_{j \in N_i(\tilde{\mathbf{x}}^{\triangleright i}(k))} \tilde{v}_j(k) \right) - v_i(k) \right) \quad (10)$$

(2) *Cohesion*: agents move towards the centroid of the agents in the local neighborhood.

$$a_i^c(k) = w_c \cdot \left( \left( \frac{1}{|N_i(\tilde{\mathbf{x}}^{\triangleright i}(k))|} \cdot \sum_{j \in N_i(\tilde{\mathbf{x}}^{\triangleright i}(k))} \tilde{x}_j(k) \right) - x_i(k) \right) \quad (11)$$

(3) *Separation*: agents move away from nearby neighbors.

$$a_i^s(k) = w_s \cdot \frac{1}{|N_i(\tilde{\mathbf{x}}^{\triangleright i}(k))|} \cdot \left( \sum_{j \in N_i(\tilde{\mathbf{x}}^{\triangleright i}(k))} \frac{x_i(k) - \tilde{x}_j(k)}{\|x_i(k) - \tilde{x}_j(k)\|^2} \right) \quad (12)$$

The cohesion and alignment rules help form and maintain a closely packed, flock-like formation. The separation rule prevents agents from coming too close to each other, thus reducing crowding and collisions.

Non-negative constants $w_{al}$, $w_c$ and $w_s$ are the weights for each acceleration component. Typically, a smaller interaction radius (hence a smaller neighborhood) is used for the separation rule, because it is significant only when agents are very close to each other. The overall acceleration in Reynolds' model is given by:

$$a_i(k) = a_i^{al}(k) + a_i^c(k) + a_i^s(k). \quad (13)$$

## 2.3 Olfati-Saber's potential-based model

In potential-based flocking models, the interaction between a pair of agents is modeled by a potential field. It is assumed that an agent is a point source, and it has a potential field around it, which exerts a force, equal to its gradient, on other agents in its range of influence. The potential field has circular symmetry and hence is a function of distance from the source. In the work of Olfati-Saber [7], the potential function $\psi_\alpha$ for a pair of agents has its minimum at the desired inter-agent distance $d$ of the desired $\alpha$-lattice. Outside the interaction radius $r$, the potential function is constant, so the potential field exerts no force. The exact definition of $\psi_\alpha$ is complicated: it is the definite integral of an "action function" $\phi_\alpha$ that is the product of a "bump function" $\rho_h$ and an uneven sigmoidal function $\phi$. The control law computes an agent's acceleration based on the sum of the forces from all other agents in its neighborhood and a velocity alignment term.

## 2.4 MPC-based models

Model predictive control (MPC) [2] is a well-established control technique that works as follows: at each time step $k$, it computes the optimal control sequence (agents' accelerations in our case) that minimizes a given cost function with respect to a predictive model of the controlled system and a finite prediction horizon of length $T$, i.e., from step $k + 1$ to $k + T$. Then, the first control input of the optimal sequence is applied (the remainder of the sequence is unused), and the algorithm proceeds with a new iteration.

Two main kinds of MPC-based flocking models exist, *centralized* and *distributed*. Centralized models assume that information about positions and velocities of all agents is available to compute their optimal accelerations. Formally, at each time step $k$, it solves the following optimization problem:

$$\min_{\mathbf{a}(k|k), \ldots, \mathbf{a}(k+T-1|k) \in A} J(k) + \lambda \cdot \sum_{t=0}^{T-1} \|\mathbf{a}(k + t \mid k)\|^2 \quad (14)$$

where $\mathbf{a}(k + t \mid k)$ is the control input (accelerations) for all agents at predicted time step $k + t$ starting from step $k$. The first term $J(k)$ is the primary model-specific cost function that the controller seeks to optimize within the prediction horizon; it is implicitly a function of the predicted configurations during the prediction horizon for time step $k$. The second term is standard for MPC problems and penalizes large control inputs, with weight $\lambda > 0$.

In distributed flocking models, each agent computes its optimal acceleration based only on information about its neighbors. Each agent $i$ solves an optimization problem of the form:

$$\min_{a_i(k|k), \ldots, a_i(k+T-1|k) \in A} J_i(k) + \lambda \cdot \sum_{t=0}^{T-1} \|a_i(k + t \mid k)\|^2 \quad (15)$$

where $a_i(k + t \mid k)$ is the acceleration for agent $i$ at predicted time step $k + t$ starting from step $k$, and $J_i(k)$ is the model-specific cost function for agent $i$. In distributed MPC, an agent has no way to know current or future control decisions of its neighbors, which are needed to make accurate predictions about their behavior. To address this problem, some approaches allow agents to communicate their local control decisions or future positions (e.g. [16, 18]), or assume that neighbors follow some default motion law, e.g., they move with constant velocities. We adopt the second strategy, because it does not require any communication.

The majority of existing MPC-based approaches to flocking are designed to optimize the regularity of the flock, by penalizing configurations where neighboring agents are not exactly distance $d$ apart, i.e., configurations that differ from an $\alpha$-lattice [15–18]. We call these approaches *lattice-based MPC*. Next we describe representative centralized and distributed lattice-based MPC flocking models, which we extend to account for sensing noise. The centralized model is a variant of a model by Zhan and Li [15, 16]. The distributed model is by Zhang et al. [17].

*2.4.1 Centralized lattice-based MPC flocking.* The centralized lattice-based MPC problem is defined as:

$$\min_{\mathbf{a}(k|k), \ldots, \mathbf{a}(k+T-1|k) \in A} \sum_{t=1}^{T} \|g(\mathbf{x}(k + t \mid k))\|^2 + \lambda \cdot \sum_{t=0}^{T-1} \|\mathbf{a}(k+t \mid k)\|^2 \quad (16)$$

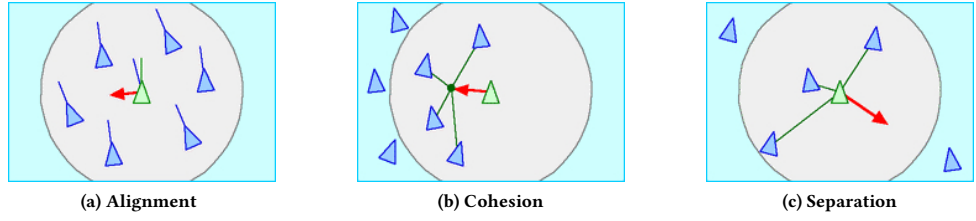

(a) Alignment  (b) Cohesion  (c) Separation

Figure 2: Interaction rules for flocking behavior in Reynolds Model.

where $\mathbf{x}(k + t \mid k)$ is the configuration of the system at predicted time step $k + t$ starting from step $k$, following the dynamics:

$$x_i(k \mid k) = \tilde{x}_i(k) \quad v_i(k \mid k) = \tilde{v}_i(k)$$
$$x_i(k + t + 1 \mid k) = x_i(k + t \mid k) + dT \cdot v_i(k + t \mid k)$$
$$v_i(k + t + 1 \mid k) = v_i(k + t \mid k) + dT \cdot a_i(k + t \mid k),$$

where the initial state of the prediction window is given by noisy measurements. For configuration $\mathbf{x}$, $g(\mathbf{x})$ captures the $\alpha$-lattice irregularity as the total deviation between agent distances and $d$:

$$\|g(\mathbf{x})\|^2 = \sum_{(i,j) \in \mathcal{E}(\mathbf{x})} \left\| x_{ji} - \frac{d \cdot x_{ji}}{\|x_{ji}\|} \right\|^2, \text{ with } x_{ji} = x_j - x_i. \quad (17)$$

This model is inspired by [15] and [16] but differs from both: in [15] the cost function also contains a velocity alignment term, which the same authors removed in their subsequent work, while in [16], "impulsive MPC" is used, which means that agents directly control their velocities (instead of accelerations), an abstraction that allows physically unrealizable accelerations.

*2.4.2 Distributed lattice-based MPC flocking.* In the distributed MPC flocking model of Zhang et al. [17], each agent $i$ controls its acceleration based on position and velocity measurements of the neighbors and assumes they have constant velocity (zero acceleration) during the prediction horizon. Similarly, the set of neighbors of $i$ is assumed invariant during the prediction horizon, and we denote it by $N_i(k) = N_i(\mathbf{x}(k))$. The control law for agent $i$ is:

$$\min_{a_i(k|k),\ldots,a_i(k+T-1|k) \in A} \sum_{t=1}^{T} \|g_i(\mathbf{x}(k + t \mid k))\|^2 +$$
$$\lambda \cdot \sum_{t=0}^{T-1} \|a_i(k + t \mid k)\|^2. \quad (18)$$

where the predicted future dynamics of $i$ is determined by:

$$x_i(k \mid k) = \tilde{x}_i(k) \quad v_i(k \mid k) = \tilde{v}_i(k)$$
$$x_i(k + t + 1 \mid k) = x_i(k + t \mid k) + dT \cdot v_i(k + t \mid k)$$
$$x_i(k + t + 1 \mid k) = x_i(k + t \mid k) + dT \cdot a_i(k + t \mid k),$$

while $i$'s neighbors $j \in N_i(k)$ have constant velocity:

$$x_j(k \mid k) = \tilde{x}_j(k) \quad x_j(k + t + 1 \mid k) = x_j(k + t \mid k) + dT \cdot \tilde{v}_j(k).$$

For configuration $\mathbf{x}$, $g_i(\mathbf{x})$ is defined in a similar way to Eq. (17) and quantifies how much $i$'s neighborhood $N_i(k)$ deviates from an $\alpha$-lattice:

$$\|g_i(\mathbf{x})\|^2 = \sum_{j \in N_i(k)} \left\| x_{ji} - \frac{d \cdot x_{ji}}{\|x_{ji}\|} \right\|^2. \quad (19)$$

## 3 DECLARATIVE FLOCKING

This section introduces centralized and distributed versions of our Declarative Flocking (DF) model, and presents a flocking algorithm based on MPC. Our formulation is declarative in that it consists of just two simple terms: (1) a cohesion term based on the average squared distance between pairs of agents, to keep the flock together, and (2) a separation term based on the inverse squared distances between pairs of agents, to avoid crowding. These two terms represent opposing forces on agents, causing agents to move towards positions in which these forces are balanced. Unlike the majority of existing MPC-based approaches that are designed to optimize conformance to an $\alpha$-lattice, our design does not impose a specific geometric structure.

### 3.1 Centralized DF model

The cost function $J$ for our centralized DF model contains the two terms described above, with the cohesion term considering all pairs of agents, and the separation term considering only pairs of agents that are neighbors. The weight $\omega$ of the separation term provides control over the density of the flock.

$$J^C(\mathbf{x}) = \frac{2}{|\mathcal{B}| \cdot (|\mathcal{B}| - 1)} \cdot \sum_{i \in \mathcal{B}} \sum_{j \in \mathcal{B}, i < j} \|x_{ij}\|^2 + \omega \cdot \sum_{(i,j) \in \mathcal{E}(\mathbf{x})} \frac{1}{\|x_{ij}\|^2}$$

The control law is Eq. (14) with $J(k)$ equal to $\sum_{t=1}^{T} J^C(\mathbf{x}(k + t \mid k))$.

### 3.2 Distributed DF model

The cost function $J$ for our distributed DF model is similar to the centralized one, except that both terms are limited to consider pairs of agents that are neighbors.

$$J_i^D(\mathbf{x}) = \frac{1}{|N_i(k)|} \cdot \sum_{j \in N_i(k)} \|x_{ij}\|^2 + \omega \cdot \sum_{j \in N_i(k)} \frac{1}{\|x_{ij}\|^2} \quad (20)$$

The control law for agent $i$ is Eq. (15) with $J_i(k)$ equal to $\sum_{t=1}^{T} J_i^D(\mathbf{x}(k + t \mid k))$.

## 4 MEASURES OF FLOCKING PERFORMANCE

We introduce four key measures of flocking performance. A single measure is insufficient, because flocking is indeed characterized by multiple desirable properties, such as aligned velocities and cohesion. Olfati-Saber introduces four main properties for flocking [7], informally described as:

(1) the group of agents stays *connected* in a unique flock, i.e., no sub-flocks and fragmentation should emerge;
(2) the group remains *cohesive*, in a close-knit formation;

(3) the group moves in a coherent way as if it was a unique body, i.e., agents' velocities are aligned; and
(4) the group maintains a regular geometry (in the $\alpha$-lattice sense).

We introduce the following four measures to capture these four requirements. An important concept in these definitions is a *sub-flock*, which is a set of interacting agents that is too far apart from other agents to interact with them. Formally, a sub-flock in a configuration $\mathbf{x}$ corresponds to a connected component of the proximity net $G(\mathbf{x})$. Let $CC(\mathbf{x}) \subseteq 2^{\mathcal{B}}$ be the set of connected components of the proximity net $G(\mathbf{x})$.

(1) The *number of connected components* of the proximity net quantifies connectedness—or, equivalently, fragmentation—of the flock. There is no fragmentation when $|CC(\mathbf{x})| = 1$. Fragmentation exists when $|CC(\mathbf{x})| > 1$. Fragmentation may be temporary or, if sub-flocks move in different directions, permanent.

(2) The *maximum component diameter*, denoted $D(\mathbf{x})$, quantifies cohesion. It is defined by

$$D(\mathbf{x}) = \max_{\mathcal{B}' \in CC(\mathbf{x})} D(\mathbf{x}, \mathcal{B}') \quad (21)$$

where $D(\mathbf{x}, \mathcal{B}')$ is the diameter of connected component $\mathcal{B}'$:

$$D(\mathbf{x}, \mathcal{B}') = \max_{\substack{(i,j) \in \mathcal{B}' \times \mathcal{B}' \\ i \neq j}} \|\mathbf{x}_{ij}\|. \quad (22)$$

Note that when all agents are isolated, i.e., $CC(\mathbf{x}) = \bigcup_{i \in \mathcal{B}} \{\{i\}\}$, $D(\mathbf{x}) = -\infty$ because the domain of the max function in Equation 22 is empty when $\mathcal{B}'$ is a singleton. Note that we consider the maximum diameter of a sub-flock in order to make this measure more independent of connectedness. If we instead considered the overall diameter of the entire (possibly fragmented) flock, any flocking model that did poorly on connectedness would also do very poorly on this measure.

(3) The *velocity convergence* measure, adopted from [17], quantifies the average discrepancy between each agent's velocity and the average velocity of the flock. In particular, we extend the measure of [17] to average velocity convergence values across sub-flocks:

$$VC(\mathbf{x}, \mathbf{v}) = \frac{\sum_{\mathcal{B}' \in CC(\mathbf{x})} \left\| \sum_{i \in \mathcal{B}'} v_i - \left(\frac{\sum_{j \in \mathcal{B}'} v_j}{|\mathcal{B}'|}\right) \right\|^2 / |\mathcal{B}'|}{|CC(\mathbf{x})|} \quad (23)$$

(4) To measure the regularity of the geometric structure of a flock, as reflected in the inter-agent spacing, we introduce a parameter-free and model-independent *irregularity* measure $I(\mathbf{x})$. For a connected component (sub-flock) $\mathcal{B}'$, it is defined as the sample standard deviation of the distances between each agent in $\mathcal{B}'$ and its closest neighbor. Thus, the measure penalizes configurations where there is dispersion in inter-agent distances, while not imposing any fixed distance between them (unlike $\alpha$-lattices).

Let $CC'(\mathbf{x}) = CC(\mathbf{x}) \setminus \bigcup_{i \in \mathcal{B}} \{\{i\}\}$ be the set of connected components where isolated agents are excluded. For $|CC'(\mathbf{x})| = 0$ (or equivalently, $|CC(\mathbf{x})| = |\mathcal{B}|$), i.e., all agents are isolated, we set the irregularity $I(\mathbf{x}) = 0$, which is the optimal value. This reflects the fact that a single point is a regular structure on its own. Moreover, such a configuration is already highly penalized by $|CC(\mathbf{x})|$ and $VC(\mathbf{v})$. For $|CC'(\mathbf{x})| > 0$, the measure is defined by:

$$I(\mathbf{x}) = \frac{\sum_{\mathcal{B}' \in CC'} \sigma\left(\biguplus_{i \in \mathcal{B}'} \min_{j \neq i} \|x_{ij}\|\right)}{|CC'|}. \quad (24)$$

where $\sigma(S)$ is the standard deviation of the multiset of samples $S$ and $\biguplus$ is the sum operator (or disjoint union) for multisets.

An $\alpha$-lattice (see Def. 2.1) has the optimal value of $I(\mathbf{x})$, i.e., $I(\mathbf{x}) = 0$, since all neighboring agents are located at the same distance $d$ from each other, leading to zero standard deviation for the term $\sigma(\{d, d, \ldots, d\})$. This shows that $I(\mathbf{x})$ captures the regularity underlying the concept of $\alpha$-lattice.

We introduce this measure because previous measures of regularity or irregularity, such as those in [7, 16, 17], measure deviations from an $\alpha$-lattice with a specified inter-agent distance $d$ and are therefore inapplicable to flocking models, such as Reynolds' model and our DF models, that are not based on $\alpha$-lattices and do not have a specified target inter-agent distance. Also, our irregularity measure is more flexible than those based on $\alpha$-lattices, because it gives an optimal score to some configurations that are geometrically regular but not $\alpha$-lattices. For example, consider a configuration $\mathbf{x}$ in which the agents are on the vertices of a grid with edge length $e$, and the interaction radius is equal to the length of the diagonal of a box in the grid. This configuration has an optimal value for our irregularity measure, i.e., $I(\mathbf{x}) = 0$, because the distance from every agent to its nearest neighbor is $e$. This configuration is not an $\alpha$-lattice and hence does not nave an optimal value for the irregularity measures used in prior work.

## 5 PERFORMANCE EVALUATION

We compare the performance of the models of Section 2 with the newly introduced DF flocking models in the 2-dimensional setting. In the first set of experiments (Section 5.1), we evaluate the performance measures illustrated in Section 4. In the second set of experiments (Section 5.2), we analyze the resilience of the algorithms to sensor noise.

For consistency with the experimental settings of [17], the lattice-based MPC problems are solved using the interior point method implemented in MATLAB's fmincon function. Our DF-MPC problems are solved using gradient descent optimization. Unless otherwise specified, the population size is $n = 30$, the simulation length is 100, $dT = 0.3$, $\bar{v} = 8$, $\bar{a} = 1$, $r = 8.4$, $d = 7$, $T = 3$, and $\lambda = 1$. These parameter values are the same ones reported in [17]. Following the settings in the OpenSteer project [11], the parameters for Reynolds' model are $r_c = 9$, $r_s = 5$, $r_{al} = 7.5$, $w_c = 8$, $w_s = 12$, and $w_{al} = 8$. The weight of the separation term in our centralized and distributed DF-MPC is $\omega = 50$. As in [17], initial positions and initial velocities of agents are uniformly sampled from $[-15, 15]^2$ and $[0, 2]^2$, respectively.

### 5.1 Performance Comparison of Flocking Algorithms

Fig. 3 shows examples of final formations for all flocking models. In particular, we chose configurations where fragmentation did not occur. We observe that the formations for lattice-based MPC algorithms have spread-out, rigid structures, consistent with the design objective of maximizing the $\alpha$-lattice regularity. On the other hand, Reynolds and our DF MPC models result in more natural flock shapes.

In Fig. 4, we compare the performance measures averaged over 100 runs for each flocking model. Regarding the number of connected components (sub-flocks), our centralized DF-MPC registers the best behavior, rapidly stabilizing to an average of 1 component (see plot a). Our distributed DF-MPC and Reynolds' model have comparable performance, reaching an average number of sub-flocks below 1.4. The lattice-based MPCs and Olfati-Saber instead lead to constant fragmentation, with more than 2 sub-flocks for the distributed lattice-based MPC, 6 for the centralized lattice-based MPC, and more than 8 for Olfati-Saber's model.

This ranking is confirmed by the diameter measure (plot b), where our centralized and distributed DF-MPC and Reynolds' model show the best cohesion, outperforming the lattice-based approaches. Recall that this measure indicates the maximum diameter over all sub-flocks, not the diameter of the entire population. As a consequence, fragmentation tends to improve diameter values since it produces sub-flocks with fewer individuals. This explains why our distributed DF-MPC performs better on this measure than the centralized version, and similarly why Olfati-Saber's model has smaller diameter measure than centralized lattice-based MPC, which in turn has smaller diameter measure than the distributed variant.

As expected, Olfati-Saber's model and the lattice-based MPCs have very good performance for irregularity (plot c), since they are designed to achieve the regular geometric formation of $\alpha$-lattice. Surprisingly, our distributed DF-MPC performs almost as well as them on this measure. Centralized DF-MPC and Reynolds' model have the least regular formations.

For velocity convergence (plot d), we find that all models perform comparably well and are able to achieve flocks with consistent velocities fairly quickly after an initial spike.

### 5.2 Robustness to Sensing Noise

To evaluate the resiliency of the models to sensor noise, we performed 20 runs for each model at 10 noise levels. The noise levels are numbered from 1 to 10, and noise level $i$ has $\sigma_x = 0.2i$ and $\sigma_v = 0.1i$. For each performance metric, we averaged its final values over 20 runs for each noise level. The results are plotted in Fig. 5. Of the six models, Olfati-Saber's model is the most vulnerable to sensing noise: the number of sub-flocks $|CC|$ in Olfati-Saber's model quickly increases to nearly 30, rendering other metrics irrelevant. The lattice-based MPC models also exhibit high fragmentation, leading to nominally good but largely irrelevant values for the other performance metrics. Our distributed DF-MPC and Reynolds' model have the best resiliency to sensing noise, with both models exhibiting similar profiles in all metrics. While the irregularity and velocity convergence measures increase with noise level, as expected, both models remarkably maintain almost a single connected component with a nearly constant component diameter for all 10 noise levels, with DF-MPC achieving a smaller diameter than Reynolds' model.

## 6 RELATED WORK

Reynolds [12] introduced the first rule-based approach for simulation of flocking behavior. With only three simple rules, his model is able to capture complex flocking behaviors of animals. Additional rules can be added to the model to simulate specific behaviors, such as leader following and predator avoidance. Pearce *et al.* [8] present a rule-based strategy for flocking, where agents move to maximize their view out of the flock. Cucker and Dong [3] present a rule-based flocking approach with proofs of convergence and collision avoidance.

Cucker and Smale [4] introduced another popular rule-based flocking model. The Cucker-Smale model is parameterized by a constant $\beta$ such that if $\beta < 1/2$, velocity convergence is guaranteed. If $\beta \geq 1/2$, velocity convergence can also be achieved under some conditions on the initial positions and initial velocities of the agents. Ahn and Ha [1] investigated the effects of multiplicative noise on the long term dynamics of the Cucker-Smale model. Erban *et al.* [5] extend the Cucker-Smale model to take into account stochasticity (imperfections) of agent behavior and delay in agents' responses to changes in their environment.

Flocking models based on potential fields have been proposed in several papers. Tanner *et al.* [14] propose a potential function $U_{ij}$, given in Equation 25, where $||r_{ij}||^2$ is the distance between agents $i$ and $j$. For distances greater than $R$, the potential is set to a constant value, $C$, indicating a zero force. In their control law, the acceleration of agent $i$ is based on the sum over all neighbors $j$ of the gradient of the potential function $U_{ij}$.

$$U_{ij} = \begin{cases} \frac{1}{||r_{ij}||^2} + \log||r_{ij}||^2, & ||r_{ij}||^2 < R \\ C, & ||r_{ij}||^2 \geq R \end{cases} \quad (25)$$

A similar potential function is also proposed by [10]. Furthermore, potential-based solutions have been extended with additional behaviors such as obstacle avoidance and leader following. For example, Ogren et.al. [9] use the motion of the leader to guide the motion of the flock; the leader's motion is independent, i.e., is not influenced by other agents.

La and Sheng [6] propose an extension of Olfati-Saber's model designed for noisy environments. In addition to the terms found in Olfati-Saber's model, their control law contains feedback terms for position and velocity, to make agents tend to stay close to the centroid of their neighborhood and minimizing the velocity mismatch with their neighbors. They show that adding these feedback terms to the control law helps bound the error dynamics of the system.

## 7 CONCLUSIONS

This paper presents an abstract declarative form of control for flocking behavior and the results of a thorough comparison of centralized and distributed versions of our MPC-based declarative flocking with four other flocking models. Our simulation results demonstrate that DF-MPC yields the best cohesion and least fragmentation, and produces natural flock shapes like those produced by Reynolds' rule-based model. Our resiliency analysis shows that the distributed version of our DF-MPC is highly robust to sensor noise.

As future work, we plan to study resilience of the flocking models with respect to additional noisy scenarios such as actuation noise (i.e., noise affecting acceleration) and faulty agents with deviant behavior. We also plan to investigate smoothing techniques to increase resilience to sensor noise.

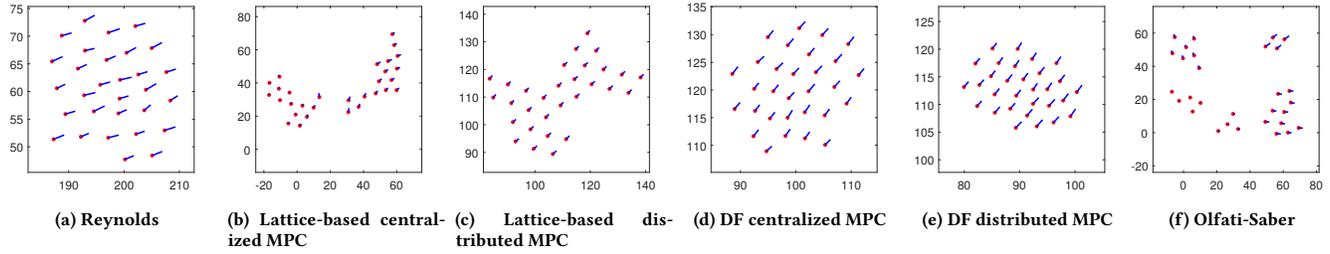

**(a) Reynolds**　　**(b) Lattice-based centralized MPC**　　**(c) Lattice-based distributed MPC**　　**(d) DF centralized MPC**　　**(e) DF distributed MPC**　　**(f) Olfati-Saber**

**Figure 3: Examples of final formations for different flocking models. The red dots are the agent positions. The blue lines denote the agent velocities; the line lengths are proportional to the speeds.**

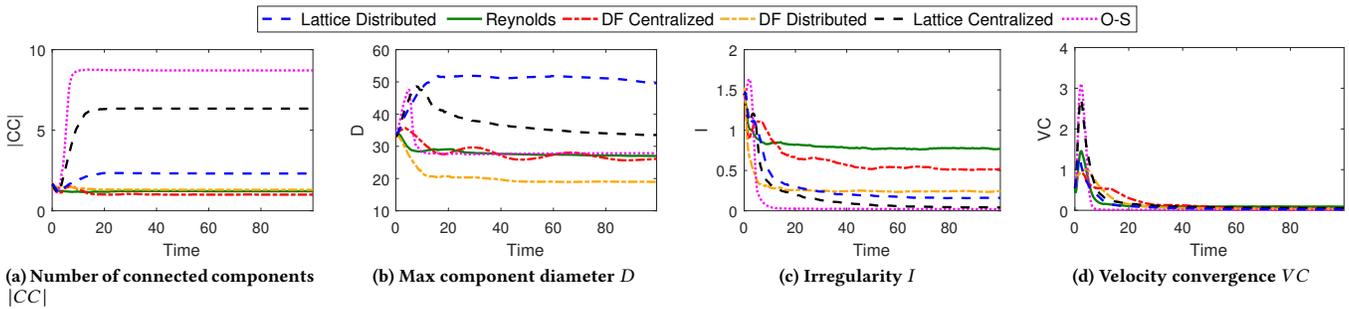

**(a) Number of connected components $|CC|$**　　**(b) Max component diameter $D$**　　**(c) Irregularity $I$**　　**(d) Velocity convergence $VC$**

**Figure 4: Comparison of performance measures obtained with 100 runs for each flocking algorithm.**

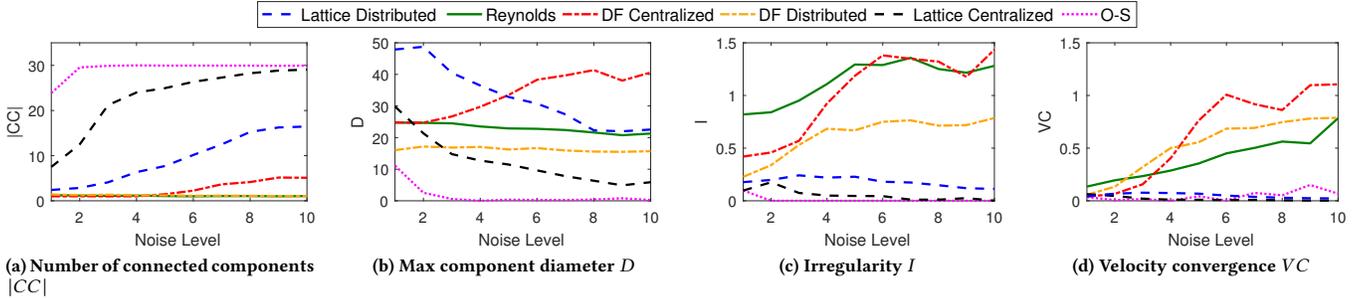

**(a) Number of connected components $|CC|$**　　**(b) Max component diameter $D$**　　**(c) Irregularity $I$**　　**(d) Velocity convergence $VC$**

**Figure 5: Comparison of the final values of the performance measures obtained with 20 runs for each flocking algorithm and for each noise level.**